\documentclass[twocolumn,showpacs,preprintnumbers,amsmath,amssymb,epsf]{revtex4}
\usepackage{graphicx}% Include figure files
\usepackage{dcolumn}% Align table columns on decimal point
\usepackage{bm}% bold math
\begin{document}
%\preprint{APS/123-QED}
\title{Signatures of doubly charged Higgs in SU(3)$_{L}\otimes$U(1)$_{N}$ model}   
\author{J. E. Cieza Montalvo$^1$, Nelson V. Cortez Jr.$^2$, M. D. Tonasse$^3$}   
\address{$^1$Instituto de F\'{\i}sica, Universidade do Estado do Rio de Janeiro, Rua S\~ao Francisco Xavier 524, 20559-900 Rio de Janeiro, RJ, Brazil}
\address{$^2$Justino Boschetti 40, 02205-050 S\~ao Paulo, SP, Brazil}
\address{$^3$Unidade de Registro, {\it Campus} Experimental de Registro, Universidade Estadual Paulista, Rua Tamekishi Takano 5, 11900-000, Registro, SP, Brazil}   
\date{\today}
%Lines break
%automatically or can be forced with \\
% It is always\today, today,
% but any date may be explicitly specified
\begin{abstract}
The scalar sector of the simplest version of the 3-3-1 electroweak models is constructed with three Higgs triplets only. We show that a relation involving two of the constants of the Higgs potential, two vacuum expectation values of the neutral scalars and the mass of the doubly charged Higgs boson leads to important information concerning the signals of this scalar particle. 
\end{abstract}
\maketitle
Doubly charged Higgs bosons are a kind of scalar bosons present in electroweak models which predict scalar triplets with non-zero hipercharge \cite{BAea82}. In the left-right symmetric models, for instance, they are fundamental ingredients for implementation of the seesaw mechanism to generation of neutrino masses  \cite{pati}.\par
Another important class of electroweak models which predict doubly charged Higgs bosons is the called 3-3-1 models \cite{PP92,PT93a}. Here we work with the model version of the Ref. \cite{PT93a}. In the scalar sector we have  only the three scalar triplets $\eta = \left(\begin{array}{ccc} \eta^0 & \eta^-_1 & \eta^+_2\end{array}\right)^{\tt T}$, $\rho = \left(\begin{array}{ccc} \rho^- & \rho^0 & \rho^{++}\end{array}\right)^{\tt T}$ and $\chi = \left(\begin{array}{ccc} \chi^- & \chi^{--} & \chi^0\end{array}\right)^{\tt T}$, transforming as $\left({\bf 3}, 0\right)$, $\left({\bf 3}, 1\right)$ and $\left({\bf 3}, -1\right)$, respectively, and which are necessary and sufficient to breakdown the symmetry and generate the correct mass spectrum. The most general invariant and renormalizable Higgs potential, obeying $B - L$ discrete symmetry \cite{PT93b}, is
\begin{widetext}
\begin{eqnarray}
V\left(\eta, \rho, \chi\right) & = & \mu_1^2\eta^\dagger\eta + \mu_2^2\rho^\dagger\rho + \mu_3^2\chi^\dagger\chi + \lambda_1\left(\eta^\dagger\eta\right)^2 + \lambda_2\left(\rho^\dagger\rho\right)^2 + \lambda_3\left(\chi^\dagger\chi\right)^2 + \cr
&& + \left(\eta^\dagger\eta\right)\left[\lambda_4\left(\rho^\dagger\rho\right) + \lambda_5\left(\chi^\dagger\chi\right)\right] + \lambda_6\left(\rho^\dagger\rho\right)\left(\chi^\dagger\chi\right) + \lambda_7\left(\rho^\dagger\eta\right)\left(\eta^\dagger\rho\right) + \cr 
&& +  \lambda_8\left(\chi^\dagger\eta\right)\left(\eta^\dagger\chi\right) + \lambda_9\left(\rho^\dagger\chi\right)\left(\chi^\dagger\rho\right) + \frac{1}{2}\left(f\epsilon^{ijk}\eta_i\rho_j\chi_k + {\mbox{H. c.}}\right)
\label{pot}
\end{eqnarray} 
\end{widetext}
In the potential (\ref{pot}) $f$ and $\mu_j$ $\left(j = 1, 2, 3\right)$ are  constants with dimension of mass and the $\lambda_i$ $\left(i = 1, \dots, 9\right)$ are adimensional constants \cite{TO96}.\par
The neutral components of the scalars triplets $\eta$, $\rho$ and $\chi$ develop non zero vacuum expectation values $\langle\eta^0\rangle = v_\eta$, $\langle\rho^0\rangle = v_\rho$ and $\langle\chi^0\rangle = v_\chi$, with $v_\eta^2 + v_\rho^2 = v_W^2 = (246 \mbox{ GeV})^2$ . The pattern of symmetry breaking is $\mbox{SU(3)}_L \otimes\mbox{U(1)}_N\stackrel{\langle\chi\rangle}{\longmapsto} \mbox{SU(2)}_L\otimes\mbox{U(1)}_Y\stackrel{\langle\eta, \rho\rangle}{\longmapsto}\mbox{U(1)}_{\rm em}$. From the Higgs potential (\ref{pot}) we can find the masses of the of the neutral scalars
\begin{subequations}
\begin{eqnarray}
& m^2_{H_1^0} \approx 4\frac{\lambda_2v_\rho^4 - 2\lambda_1v_\eta^4}{v_\eta^2 - v_\rho^2}, \quad m_{H_2^0}^2 \approx \frac{v_W^2v_\chi^2}{2v_\eta v_\rho}, \label{2a}\\ 
& m_{H_3^0}^2 \approx -4\lambda_3v_\chi^2, \quad m_h^2 = -\frac{fv_\chi}{v_\eta v_\rho}\left[v_W^2 + \left(\frac{v_\eta v_\rho}{v_\chi}\right)^2\right]. \label{2b}
\end{eqnarray}
It should be notice that the Eqs. (\ref{2b}) impose $\lambda_3 < 0$ and $f < 0$. The aproximations in Eqs. (\ref{2a}) and (\ref{2b}) are valid for $v_\eta, v_\rho \ll v_\chi$. For the singly charged scalars we have
\begin{eqnarray}
m_{H_1^\pm} = \frac{v_W^2}{2v_\eta v_\rho}\left(fv_\chi - 2\lambda_7v_\eta v_\rho\right), \\
\qquad
m_{H_2^\pm} = \frac{v_\eta^2 + v_\chi^2}{2v_\eta v_\chi}\left(fv_\rho - 2\lambda_8v_\eta v_\chi\right)
\end{eqnarray}
and for the doubly charged
\begin{equation}
m_{H^{\pm\pm}}^2 = \frac{v_\rho^2 + v_\chi^2}{2v_\rho v_\chi}\left(fv_\eta - 2\lambda_9v_\rho v_\chi\right).
\label{m++}
\end{equation} 
\end{subequations}
The gauge sector of the 3-3-1 model of the Ref. \cite{PT93a} acommodates one extra $Z^\prime$ neutral boson, two singly charged $(V^\pm)$ and two doubly charged $(U^{\pm\pm})$ bosons beyond the standards $Z$ and $W^\pm$ with mass spectrum 
\begin{eqnarray}
m_{Z'}^2 = \frac{2}{3}\left(\frac{e}{s_W}\right)^2(1+ 3 t^{2}) v_{\chi}^{2} ,\\
 \qquad m_V^2 = \left(\frac{e}{s_W}\right)^2\frac{v_\eta^2 + v_\chi^2}{2},\\
 \qquad m_U^2 = \left(\frac{e}{s_W}\right)^2\frac{v_\rho^2 + v_\chi^2}{2},
\end{eqnarray} 
where $s_W = \sin{\theta_W}$. We have also one heavy lepton $(E^+, M^+, T^+)$ and one heavy quark $(J_1, J_2, J_3)$ in each generation, respectively. The heavy fermions have masses of the order of $v_\chi$. For details on this version of the 3-3-1 models see Ref. \cite{PT93a}. \par
In this work we wish to report about the signals of the doubly charged Higgs when we consider particular values for the adimensional parameter $\lambda_9$ in the Higgs potential (\ref{pot}) and for the vacuum expectation value $v_\chi$ \cite{TO96}. The production  and signals for  the doubly charged Higgs bosons of the 3-3-1 models at LHC collider has been investigated in Ref. \cite{cieco1}, where has been found no good signatures for them. However, in this present work we have shown that these signatures are very significant for particular parameter values and due that we are working with masses of order of 500 GeV to 700 GeV, which give better and satisfactory results than that of the previous work. Although we have a reasonable number of free parameters [see Eqs. (\ref{pot}) to (\ref{2b})], the model establishes strong bounds among them, these constraints and bounds are really important, bringing good results as can be seen in this manuscript. Throughout this work we take the values of standard model parameters, $m_Z = 91.1876$ GeV, $s_W^2 = 0.23122$, and $m_W = 80.403$ GeV as seen in Ref. \cite{Yea06}. For the 3-3-1 model we take two parameters sets listed in the Table I.
\begin{center}
\begin{table}[h]
\caption{\label{table:1}\footnotesize\baselineskip = 12pt The two parameters sets used in this work. Values of the other variables are given throughout the text.}
\begin{tabular}{cccccccc}  
\hline\hline
Set & $\lambda_9$ & $v_\eta$ (GeV) & $\lambda_1$ & $\lambda_2$ & $\lambda_6$ & $\lambda_7$ & $\lambda_8$ \\ 
\hline 
1 & -0.8 & \raisebox{-1.5ex}{ 195} & \raisebox{-1.5ex}{ -1.2} & \raisebox{-1.5ex}{ -1} & \raisebox{-1.5ex}{ 1} & \raisebox{-1.5ex}{ -2} & \raisebox{-1.5ex}{ -1} \\
2 & -1.2 & &  &  & &  & \\
\hline\hline
\end{tabular}
\end{table}
\end{center}
We fix the general bounds of the adimensional constants of the Higgs potential (\ref{pot}) as $-3 \leq \lambda_i \leq 3$ $\left(i = 1, \ldots, 9\right)$ to guarantee approximately the pertutubative regime. Thus, from Eq. (\ref{m++}) we can see that the inequality $-3 \leq \lambda_9 \leq 0$  is satisfied from the positivity of $m_{H^{\pm\pm}}^2$ because $f \leq 0$. \\
On the other hand, if we consider $-f \approx v_\chi \gg v_\eta, v_\rho$, we have
\begin{equation}
\lambda_4 \approx 2\frac{\lambda_2v_\rho^2 - \lambda_1v_\eta^2}{v_\eta^2 - v_\rho^2}, \qquad \lambda_5v_\eta^2 + 2\lambda_6v_\rho^2 \approx -\frac{v_\eta v_\rho}{2}
\label{vin1}
\end{equation}
as seen in Ref. \cite{TO96}. \par
\begin{figure}[t]
\includegraphics [scale=0.4]{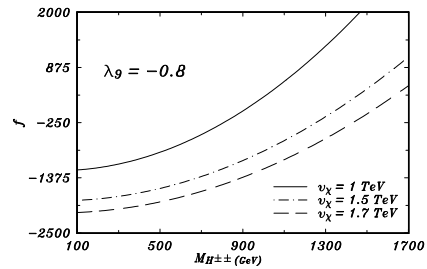}
\caption{\label{fig1} Parameter $f$ as a function of mass of doubly charged Higgs boson for $\lambda_{9}=$ -0.8.}
\end{figure}
\begin{figure}
\includegraphics [scale=0.40]{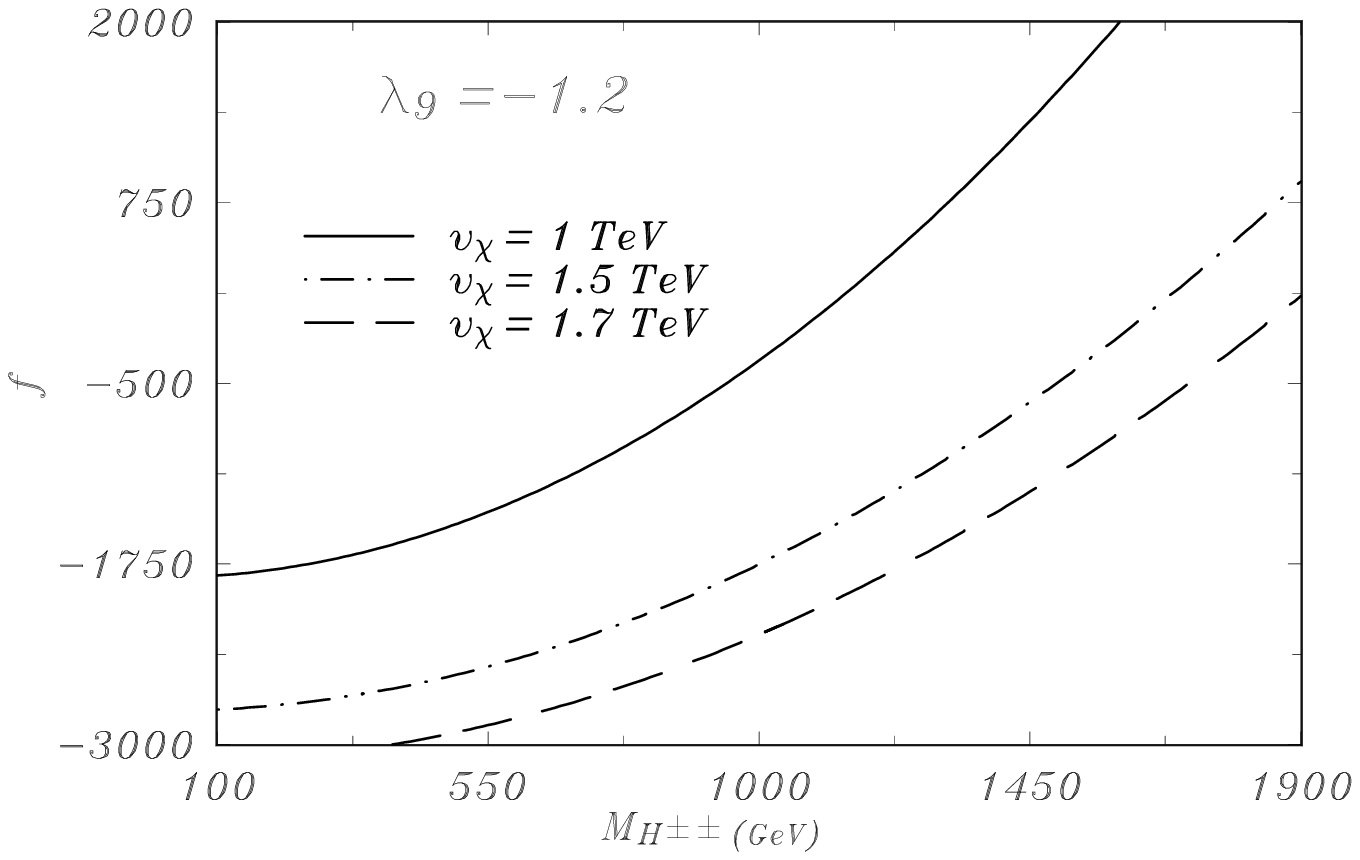}
\caption{\label{fig2} Parameter $f$ as a function of mass of doubly charged Higgs boson for $\lambda_{9}=$ -1.2.}
\end{figure}
\begin{figure}
\includegraphics [scale=0.40]{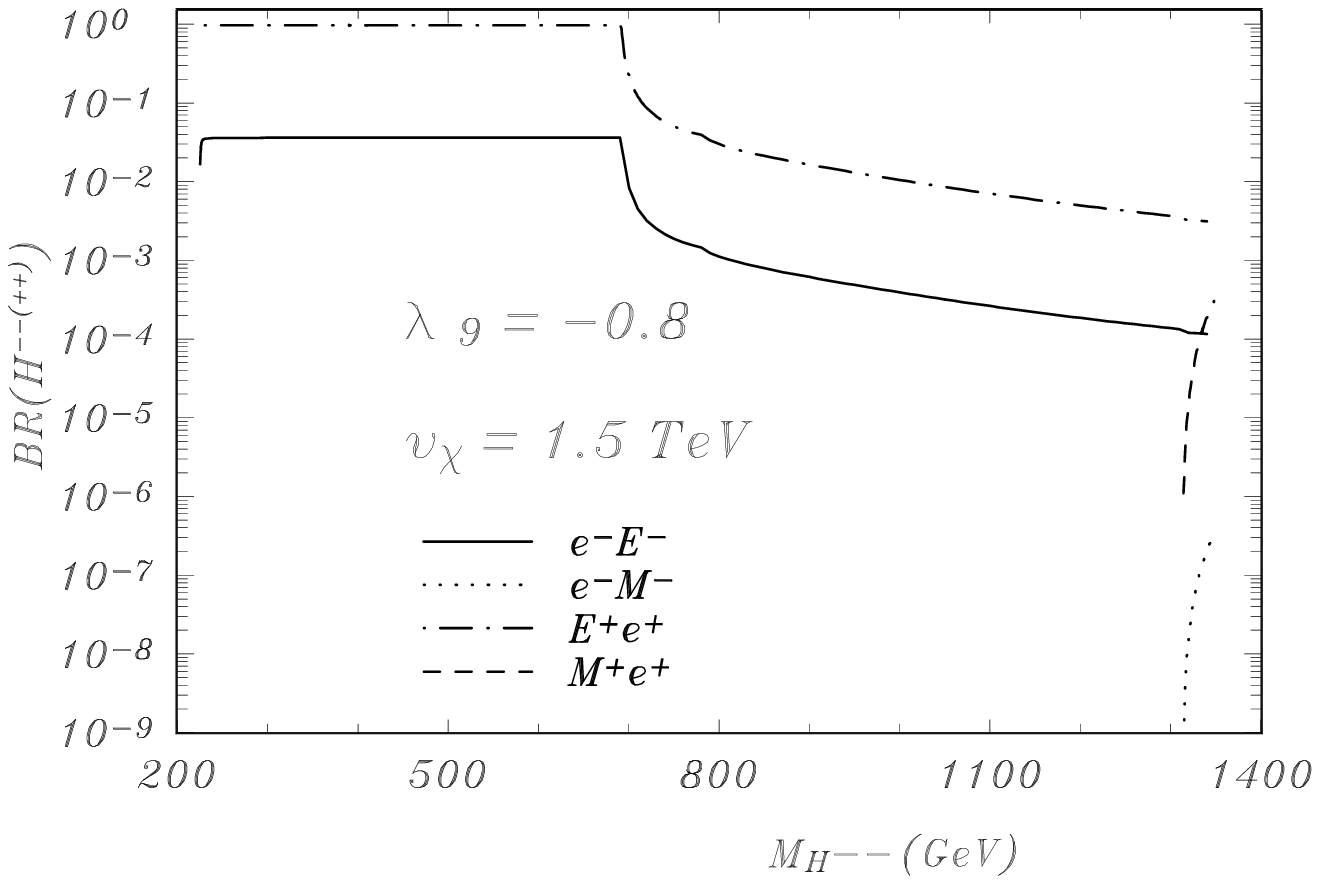}
\caption{\label{fig3}  Branching ratios for the doubly charged Higgs decays as a functions of $m_{H^{\pm \pm}}$ for $\lambda_{9}=$ -0.8 for the fermionic sector.} 
\end{figure}
\begin{figure}
\includegraphics [scale=0.40]{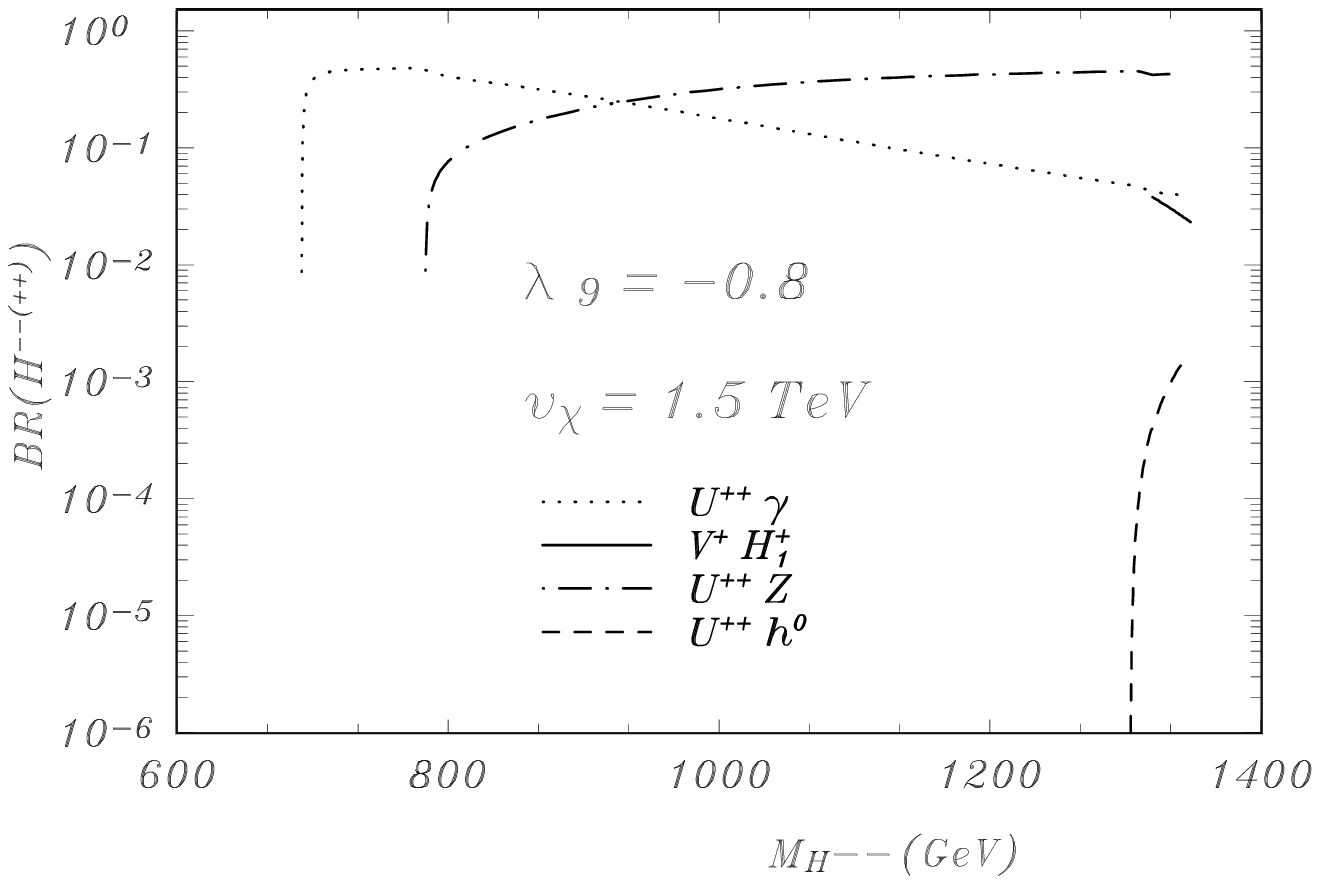}
\caption{\label{fig4} Branching ratios for the doubly charged Higgs decays as a functions of $m_{H^{\pm \pm}}$ for $\lambda_{9}=$ -0.8 for the bosonic  sector.}
\end{figure}
\begin{figure}
\includegraphics [scale=0.40]{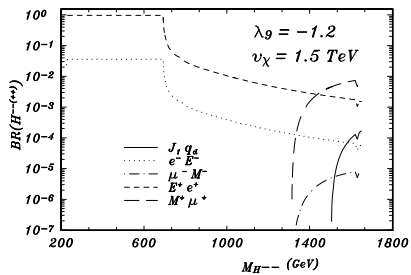}
\caption{\label{fig5} Branching ratios for the doubly charged Higgs decays as a functions of $m_{H^{\pm \pm}}$ for $\lambda_{9}=$ -1.2 for the leptonic sector.}
\end{figure}
\vskip 0.5cm
A simple analysis of Eq. (\ref{m++}) also shows that the lowest values for $\lambda_{9}$ and $f$, such to produce values for $m_{H^{\pm \pm}}$ between 100 GeV and 200 GeV, for $v_\chi$ = 1500 GeV and  $v_\rho$ = 195 GeV are $\lambda_{9}= -4.50 \times  10^{-3}$, $f= -0.23 $ GeV and $\lambda_{9}= -1.80 \times 10^{-2}$, $f= -0.92$ GeV, that is, by decreasing the value of $\lambda_{9}$ we increase the value of $m_{H^{\pm \pm}}$ with negative $f$. On the other hand, the values  for the parameters $\lambda_{5}$ and $\lambda_{6}$ lead  to the constraint $v_{\eta}>40.5$ GeV, which can be seen as following from the Eq. (\ref{vin1}). In the approximation $-f \approx v_{\chi}$ it is appropriate to choose the parameter  $-1.2 \leq \lambda_{9} \leq -0.8$  (see Table II) for the mass of $m_{H^{\pm \pm}}=500$ GeV. Fig. \ref{fig1} shows the possible values of $f$, varying the values of $m_{H^{\pm \pm}}$ for $v_{\chi}= (1000, 1500, 1700)$ GeV and considering  $\lambda_{9} =-0.8$ and $v_{\eta}=195$ GeV. We can conclude from Fig. 1 that for the $v_{\chi}=1000$ GeV we have the acceptable masses up to $m_{H^{\pm \pm}} \simeq 903$ GeV, for $v_{\chi}=1500$ GeV we will have up to $m_{H^{\pm \pm}} \simeq 1346$ GeV and for $v_{\chi}=1700$ GeV we have consequently up to $m_{H^{\pm \pm}}\simeq 1525$ GeV. If we now consider $\lambda_{9}=-1.2$ and the $v_{\chi}=1000$ GeV, from Fig. \ref{fig2} we see that the acceptable masses will be up to $m_{H^{\pm \pm}} \simeq 1107$ GeV, for $v_{\chi}=1500$ GeV we will have up to $m_{H^{\pm \pm}} \simeq 1651$ GeV and for $v_{\chi}=1700$ GeV we have consequently up to $m_{H^{\pm \pm}}\simeq 1869$ GeV. It should be noticed that the VEV $v_\eta$ and $v_\rho$ influence very little on the mass of $m_{H^{\pm\pm}}$, so for example for $\lambda_{9}= -0.5$, $v_{\chi}= 1500$ GeV and $v_{\eta}=$ 50 GeV the acceptable mass will be $m_{H^{\pm \pm}}=1074$ GeV, for $v_{\eta}=$ 150 GeV will be
  $m_{H^{\pm \pm}}=1069.5$ GeV and for $v_{\eta}=$ 240 GeV will be $m_{H^{\pm \pm}}=1061$ GeV. If we take $\lambda_{9}=-1.2$,  $v_{\chi}=1500$ GeV and $v_{\eta}=50$ GeV  then the acceptable mass will be $m_{H^{\pm \pm}}=1664$ GeV, for $v_{\eta}=150$ GeV we will have $m_{H^{\pm \pm}}=1656$ GeV  and consequently for $v_{\eta}=$ 240 GeV we have $m_{H^{\pm \pm}}=1644$ GeV,  all this happen because in the Eq.  (\ref{m++}) the $v_{\eta}$ and $v_{\rho}$ are related by the relation $v_{\eta}^{2}+ v_{\rho}^{2}= (246 \mbox{ GeV})^{2}$  and $v_{\rho}$ is smaller compared with $v_{\chi}$. \par  
\begin{widetext}
\begin{center}
\begin{table}[h]
\label{table:1}
\caption{\footnotesize\baselineskip = 12pt The parameters sets used in this work. Values of the other variables are given throughout the text.}
\begin{tabular}{l|rrrrrrrrrr}  
\hline\hline
$-f \times 10^3$ GeV & 0.25 & 0.88 & 1.50 & 2.75 & 3.99 & 0.91 & 1.62 & 2.55 & 4.42 & 6.52 \\  
$  v_\chi \times 10^3$ GeV   & 1.01 & 1.01 & 1.01 & 1.01 & 1.01 & 1.51 & 1.51 & 1.51 & 1.51 & 1.51 \\
$-\lambda_9$ &  0.40 & 0.80 & 1.20 & 2.00 & 2.80 & 0.50 & 0.80 & 1.20 & 2.00 & 2.90\\
\hline\hline
\end{tabular}
\end{table}
\end{center}
\end{widetext}

Figs. \ref{fig3} and \ref{fig4}, show the branching ratios for the Higgs decays, $H^{\pm \pm} \rightarrow {\it all}$, where we have chosen for the parameters, masses and the VEV, the following 
representative values: $\lambda_{1} =-1.2$,  $\lambda_{2}=\lambda_{3}=-\lambda_{6}=\lambda_{8}=-1$, $\lambda_{4}= 2.98$ $\lambda_{5}=-1.57$, $\lambda_{7}=-2$ and $\lambda_{9}=-0.8$, $v_{\eta}=195$ GeV and $v_{\chi}=1500$ GeV. From  Fig.  \ref{fig1} we see that the others particles, such as $J_{1,2,3}, T, U^{\pm \pm}, Z', H_{2}^{\pm}, H_{2,3}^{0}$ and $h^{0}$, do not take part in this process because of their masses are larger than the mass of the doubly charged Higgs.

\begin{figure}
\includegraphics [scale=0.40]{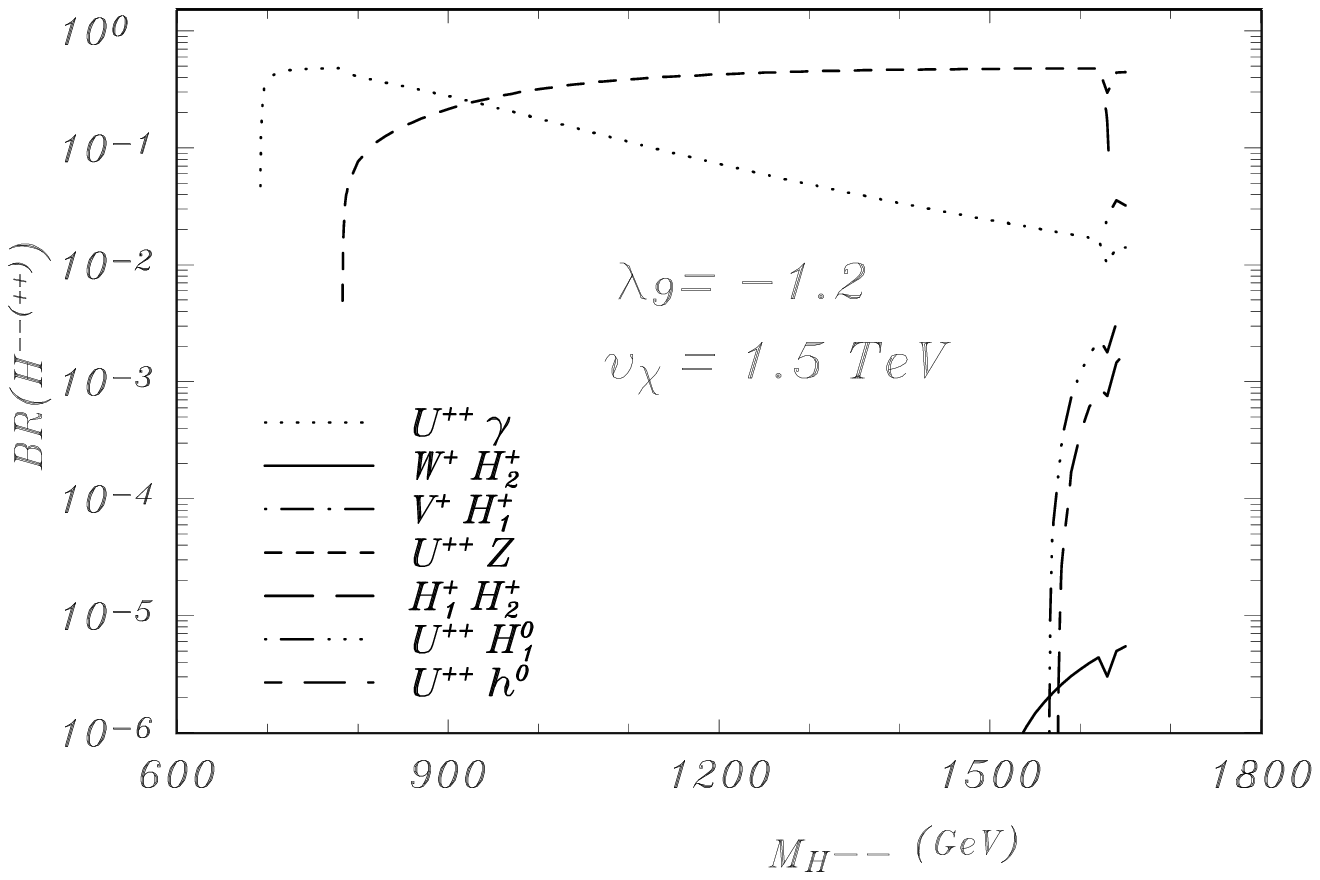}
\caption{\label{fig6}  Branching ratios for the doubly charged Higgs decays as a functions of $m_{H^{\pm \pm}}$ for $\lambda_{9}=$ -1.2 for the bosonic  sector.}
\end{figure}

\begin{figure}
\includegraphics [scale=0.40]{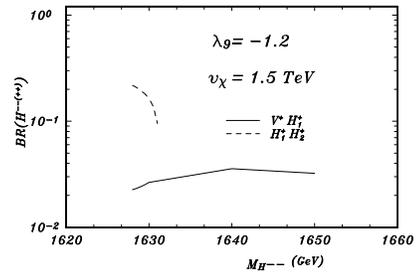}
\caption{\label{fig7} Branching ratios for the doubly charged Higgs decays as a functions of $m_{H^{\pm \pm}}$ for $\lambda_{9}=$ -1.2 for the $H^{\pm \pm} \rightarrow V^{\pm} H_{1}^{\pm}$ and $H^{\pm \pm} \rightarrow H_{1}^{\pm} H_{2}^{\pm}$.}
\end{figure}

\begin{figure}
\includegraphics [scale=0.40]{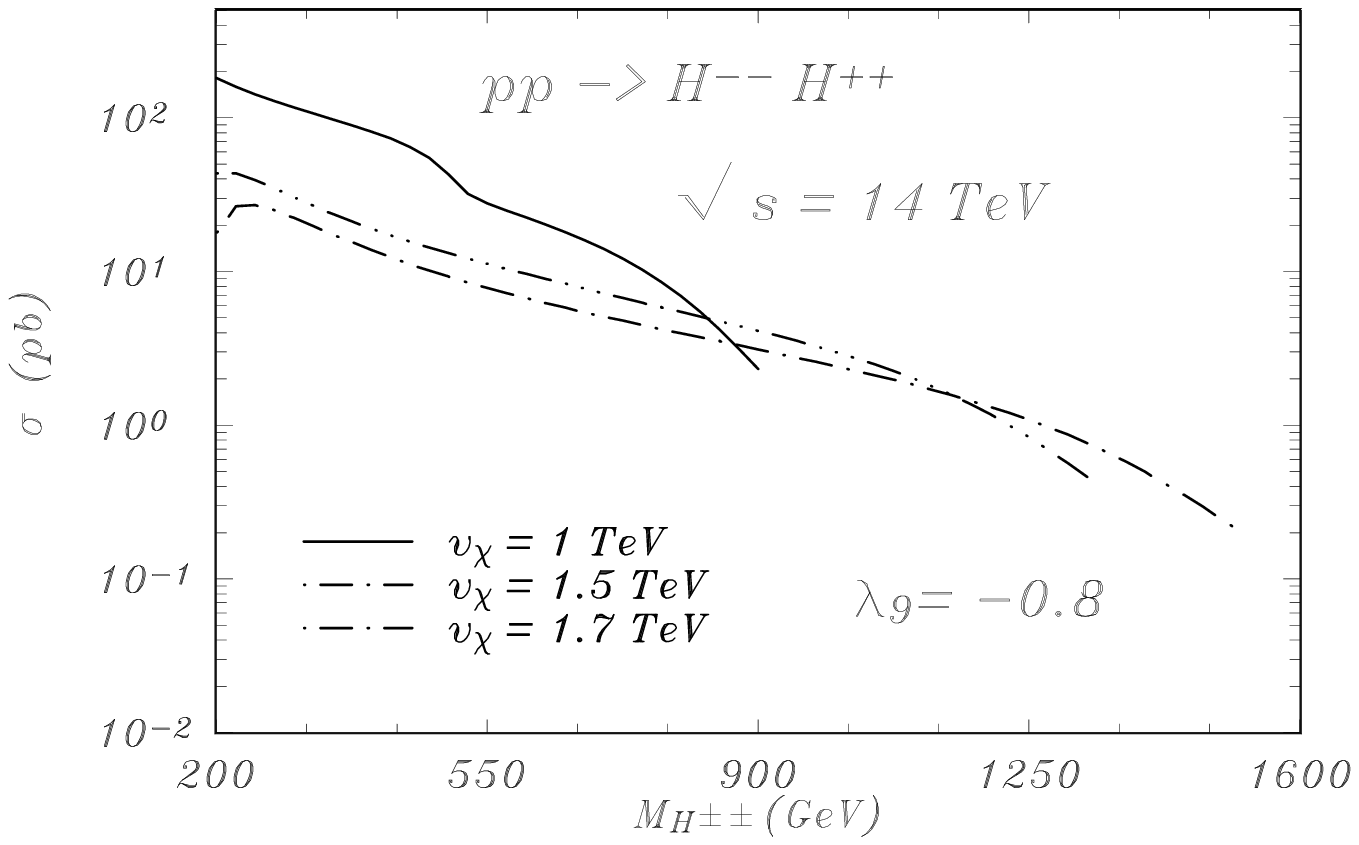}
\caption{\label{fig8}  Total cross section for the process $p p \rightarrow H^{--} H^{++}$ as a function of $m_{H^{\pm \pm}}$ for $\lambda_{9}=-0.8$ }
\end{figure}

In Figs. \ref{fig5} and \ref{fig6} we exhibit the branching ratios for the same particle and for the same parameters, masses and the VEV, as considered in Fig. \ref{fig3}, except that we consider  $\lambda_{9}= -1.2$. Considering this value, and from Fig. \ref{fig2}, we see that the acceptable mass for $m_{H^{\pm \pm}}$ is larger and consequently, more particles will take part in this process, as clearly seen from Figs. \ref{fig5} and \ref{fig6}. 
Fig. \ref{fig7} shows the branching ratios: $BR (H^{\pm \pm} \rightarrow V^{\pm} H_{1}^{\pm})$ and $BR (H^{\pm \pm} \rightarrow H_{1}^{\pm} H_{2}^{\pm})$ which vary in a short interval, 
Fig. \ref{fig8} shows the cross section for the process $pp \to H^{++}H^{--}$,  which is given in Ref. \cite{cieco1}, for the same values of $\lambda^{'}s$ considered above and $v_{\eta}=195$GeV, except which we put $\lambda_{9}=-0.8$ and with other particles masses as given in the Table II, it is to notice that the masses of $m_{h^0}$, $m_{H_1^{\pm}}$ and $m_{H_{2}^{\pm}}$  depend on the parameter $f$ and therefore they can not be fixed by any value of $v_{\chi}.$
\begin{widetext}
\begin{center}
\begin{table}[t]
\label{table:2}
\caption{\footnotesize\baselineskip = 12pt Values of the masses for $v_\eta = 195$ GeV and the sets of parameters given in the text. All the values in this table are given in GeV. For all values of the parameters for this table we have $m_{H_1^0} = 874$ GeV and $m_{H^\pm} = 348$ GeV.}
\begin{tabular}{ccccccccccccc}  
\hline\hline
$v_\chi$ & $m_E$ & $m_M$ & $m_T$ & $m_{H^0_2}$ & $m_{H^0_3}$ & $m_V$ & $m_U$ & $m_{Z^\prime}$ & $m_{J_1}$ & $m_{J_2}$ & $m_{J_3}$\\ 
\hline 
1000 & 148.9 & 875 & 2000 & 1017.2 & 2000 & 467.5 & 464 & 1707.6 & 1000 & 1410 & 1410 \\
1500 & 223.3 & 1312.5 & 3000 & 1525.8 & 3000 & 694.1 & 691.8 & 2561.3 & 1500 & 2115 & 2115 \\
1700 & 251.3 & 1487.5 & 3400 & 1729.2 & 3400 & 785.2 & 783.1 & 2902.8 & 1700 & 2397 & 2397\\
\hline\hline
\end{tabular}
\end{table}
\end{center}
\end{widetext}
We have chosen $v_{\chi}=1700$ GeV as the maximum value because it is constraint by the mass of the  $m_{Z'}=(0.5-3)$ TeV, which is proportional to $v_{\chi}$ \cite{PP92}

In our numerical analysis we have not considered the contribution of gluon-gluon fusion because it is, at least, four orders of magnitude smaller than that of Drell-Yan process. To turn this work more complete we are introducing the widht of $H^{\pm \pm} \to Uh^0$
\begin{widetext}
\begin{eqnarray}
\Gamma_{{H^{\pm \pm}} \to U^{\pm \pm}h^0} & = & \frac{AK}{32\pi m_{H^{++}}} \left(\frac{m^4_{H^{\pm \pm}}}{2m^2_{U{\pm \pm}}} - \frac{m^2_{H^{\pm \pm}}m^2_{h^{0}}}{m^2_{U{\pm \pm}}}  - m^2_{H^{\pm \pm}} + \frac{m^2_{U^{\pm \pm}}}{2} + \frac{m^4_{h^{0}}}{2m^2_{U{\pm \pm}}} - \frac{m^2_{h^{0}}}{2} \right), 
\end{eqnarray}
where
\begin{subequations}\begin{eqnarray}
A & = & \sqrt{\left[ 1 - \left( \frac{m_{U^{\pm \pm}} + m_{h^{0}}}{m_{H^{\pm \pm}}}\right)^2\right]\left[ 1 - \left( \frac{m_{U^{\pm \pm}} - m_{h^{0}}}{m_{H^{\pm \pm}}}\right)^2\right]}, \\
K & = & \frac{ieu}{2s_W\sqrt{2\left(u^2+w^2\right)}}.
\end{eqnarray}\end{subequations}
\end{widetext}

\begin{figure}
\includegraphics [scale=0.40]{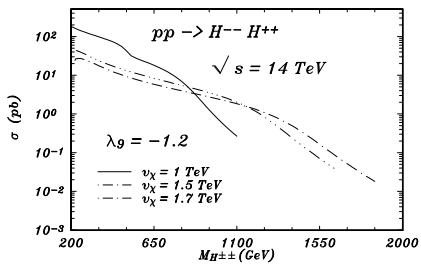}
\caption{\label{fig9} Total cross section for the process $p p \rightarrow H^{--} H^{++}$ as a function of $m_{H^{\pm \pm}}$ for $\lambda_{9}=-1.2$ }
\end{figure}

Considering that the expected integrated luminosity for the LHC will be of order of $3 \times 10^5$ pb$^{-1}$/yr then the statistics give a total of $\simeq 3.9 \times 10^{6}$ events per year for Drell-Yan process, if we take the mass of the Higgs boson $m_{H^{\pm \pm}}= 500$ GeV, $v_{\chi}=1500$ GeV, $m_{H_{2}}^{\pm}= 1163.7$ GeV, $m_h = 2229.9$ GeV and $\lambda_{9}$ = $-0.8$. Considering that the signal for $H^{\pm \pm}$ production will be $e^{-} P^{-}$ and $e^{+} P^{+}$ and taking into account that the branching ratios for both particles would be $B(H^{--} \to e^{-} P^{-}) = 3.6 \%$ and $B(H^{++} \to e^{+} P^{+}) = 96.4 \%$, see Fig. \ref{fig3}, we would have approximately $1.3 \times 10^{5}$ events per year for Drell-Yan process. Taking now the mass of the Higgs boson $m_{H^{\pm \pm}}= 700$ GeV,  $v
 _{\chi}=1500$ GeV, $m_{H_2^\pm}= 1223.6$ GeV, $m_h = 2052.2$ GeV and the same value of $\lambda_{9}$ as above. We then have a total of $\simeq 2.2 \times 10^6$ events per year for Drell-Yan process. Considering now the same signal as above, whose branching  ratios  are equal to $B(H^{--} \to e^{-} P^{-}) = 0.8 \%$ and $B(H^{++} \to e^{+} P^{+}) = 22 \%$, we will have a total of approximately $3.9 \times 10^{3}$ events for Drell-Yan process. 
In Fig. \ref{fig9} we show the cross section for the process $pp \to H^{++}H^{--}$ for the same parameter values of $\lambda^{'}s$, $v_{\eta}=195$ GeV and other particle masses considered above except for the $\lambda_{9}= -1.2$, which gives for $m_{H^{\pm \pm}}= 500$ GeV, $v_{\chi}=1500$ GeV, the values of $m_{H_{2}}^{\pm}= 901.6$ GeV  and $m_h = 2802.7$ GeV, in this case the statistics will be equal as in Fig. \ref{fig8}, because the  cross section are nearly equal and also the branching ratios, where participate the same  particles for the mass $m_{H^{\pm \pm}}=500(700)$ GeV, are also equals. So from this analysis we conclude that varying the value of $\lambda_{9}$ from $-3 <  \lambda_{9} < 0$, where in our case we choose $\lambda_{9}=-0.8$ and $\lambda_{9}=-1.2$,  and choosing the same signals, we obtain results nearly equal.   
Then we can conclude that we will have a very striking signal, the double-charged Higgs boson will deposit four times the ionization energy than the characteristic single-charged particle, that is, if we see this signal we will not only be seeing the double charged Higgs but also the heavy leptons. The main background for this signal, $p \ p \rightarrow H^{--} H^{++}  \rightarrow e^{-} P^{-} \ (e^{+} P^{+} )$, could come from the process $p \ p \rightarrow Z \ Z$ and another small background from the $p \ p \rightarrow W^{-} W^{+} Z$, all these backgrounds can be eliminated (see Ref. \cite{CP02}).\par
In summary, through this work, we have shown that in the context of the 3-3-1 model the signatures for double charged Higgs bosons can be significant in LHC collider. Our study indicates the possibility of obtaining a clear signal of these new particles with a satisfactory number of events.

\begin{widetext}
\begin{center}
\begin{table}[hbt]
\label{table:3}
\caption{\footnotesize\baselineskip = 12pt Numbers of events per year for doubly charged higgs Bosons with respect their masses and branching ratios. The total number of events per year for $m_{H^{++}}$ = $500$ GeV is $1.5$ $\times$ $10^6$ and for $m_{H^{++}}$= $700$ GeV is $1.5 \times 10^3$.}
\begin{tabular}{cccc}  
\hline\hline
{\small $BR\left(H^{--} \to e^-E^-\right)$} & ${\small BR\left(H^{++} \to e^+E^+\right)}$ & $m_{H^{++}}^2$ (GeV) & Events/yr \\ 
\hline
3.6\% & 96.4\% & 500 & $5.5 \times 10^4$ \\
0.8\% & 22\% & 700 & $1.6 \times 10^3$ \\
\hline\hline
\end{tabular}
\end{table}
\end{center}
\end{widetext}
\par

\section{ACKNOWLEDGMENTS}
One of us (JECM) would like to thank Prof. R. O. Ramos for a careful reading of the manuscript and another (N. V. C.) would like to thank the Instituto de F\' {i}sica Te\'{o}rica, of UNESP, for the use of its facilities.

\end{document}